\documentclass[twocolumn]{aastex62}
\newcommand{\lam}{$\lambda$}

\renewcommand{\ion}[2]{#1\,{\sc #2}}

\begin{document}

\title{Element abundance ratios in the quiet Sun transition region}

\author[0000-0001-9034-2925]{P.R. Young}
\affiliation{College of Science, George Mason University, 4400 University Drive, Fairfax, VA 22030, USA}
\affiliation{Code 671, NASA Goddard Space Flight Center, Greenbelt, MD 20771, USA}
\affiliation{Northumbria University, Newcastle Upon Tyne, NE1 8ST, UK}

\begin{abstract}
Element abundance ratios of magnesium to neon (Mg/Ne) and neon to oxygen (Ne/O) in the transition region of the quiet Sun have been derived by re-assessing previously published data from the Coronal Diagnostic Spectrometer on board the Solar and Heliospheric Observatory in the light of new atomic data. The quiet Sun Mg/Ne ratio is important for assessing the effect of magnetic activity on the mechanism of the first ionization potential (FIP) effect, while the Ne/O ratio can be used to infer the solar photospheric abundance of neon, which can not be measured directly. 
The average Mg/Ne ratio is found to be $0.52\pm 0.11$, which applies over the temperature region 0.2--0.7~MK, and is consistent with the earlier study.
The Ne/O ratio is, however, about 40\%\ larger, taking the value $0.24\pm 0.05$ that applies to the temperature range 0.08--0.40~MK. The  increase is mostly due to changes in ionization and recombination rates that affect the equilibrium ionization balance. If the Ne/O ratio is interpreted as reflecting the photospheric ratio, then the photospheric neon abundance is $8.08\pm 0.09$  or $8.15\pm 0.10$ (on a logarithmic scale for which hydrogen is 12), according to whether the oxygen abundances of M.~Asplund et al.\ or E.~Caffau et al.\ are used. The updated photospheric neon abundance implies a Mg/Ne FIP bias for the quiet Sun of $1.6\pm 0.6$.
\end{abstract}


\section{Introduction}

\citet{young05-fip,2005A&A...444L..45Y} used EUV spectra from the Coronal Diagnostic Spectrometer (CDS) on board the Solar and Heliospheric Observatory (SOHO) to derive element abundance ratios of magnesium to neon (Mg/Ne) and neon to oxygen (Ne/O) in the quiet Sun, and an update to the Mg/Ne results was presented by \citet{2006ESASP.617E..47Y}. 
The derived values applied to the upper transition region ($\approx$ 0.1--0.7~MK) and were unique in that they were the first to be obtained in this temperature region on the solar disk rather than at the solar limb. Quiet Sun measurements of the Mg/Ne ratio at the limb were made with  the \ion{Mg}{vi} and \ion{Ne}{vi} ions by \citet{1993ApJ...414..381F}, \citet{1999ApJ...517..516D} and \citet{2000ApJ...539L..71D}, but the latter paper noted that the temperature structure shows large changes over small distances just above the limb, which can modify the derived Mg/Ne abundance ratio. The temperature structure on the solar disk does not show such large variations with spatial position, while the availability of multiple ions for each element from the CDS spectra removes the dependence on individual ions.

The Mg/Ne ratio is important for investigating the so-called FIP effect in the solar atmosphere, the finding that elements with first ionization potentials (FIPs) $\le$ 10~eV are over-abundant compared to elements with FIPs larger than this value \citep[e.g.,][]{2000PhyS...61..222F}. The nature of the transition at this boundary is  uncertain, and \citet{schmelz12} adopted a linear transition between 9 and 11~eV.
Magnesium is a low-FIP element (7.6~eV), and neon and oxygen are high-FIP elements (21.6 and 13.6~eV, respectively). The Ne/O ratio is of interest because it potentially allows the photospheric abundance of neon to be determined if it is assumed that the atmospheric ratio represents the photospheric value (since neon and oxygen are both considered high-FIP). This is significant because the photospheric neon abundance can not be directly determined since the element has no useful absorption lines in the photospheric spectrum. The Mg/Ne FIP bias---the amount by which the ratio is enhanced or depleted compared to the photospheric value---depends on the photospheric neon abundance, hence the need to measure Ne/O.

The present work revisits the analyses of \citet{young05-fip,2005A&A...444L..45Y} using updated atomic data available in the CHIANTI atomic database \citep{2016JPhB...49g4009Y}, with a particular focus on changes in the equilibrium ionization fractions that have arisen since 2005. Firstly, ionization and recombination rate coefficients have been updated, and secondly a formula for computing the suppression of dielectronic recombination as density increases has been published \citep{nikolic13}.

The observational data used here are presented in Sect.~\ref{sect.obs}. Sect.~\ref{sect.abun} discusses the current status of photospheric abundances for oxygen, neon and magnesium, and Sect.~\ref{sect.helio} highlights the connection between abundances, the standard model of the solar interior and helioseismology measurements. Sect.~\ref{sect.ionfrac} compares the ionization equilibrium curves used in the present work with those used by \citet{young05-fip,2005A&A...444L..45Y}, and Sect.~\ref{sect.atom} gives details of other atomic data. The method for calculating the Mg/Ne and Ne/O ratios is presented in Sect.~\ref{sect.method}, and the results are given in Sect.~\ref{sect.results}. A comparison with  coronal neon and oxygen studies 
is given in Sect.~\ref{sect.cor}, and the final results are summarized in  Sect.~\ref{sect.summary}.

\section{Observations and  calibration}\label{sect.obs}

CDS consisted of a single telescope that fed two distinct spectrometers: the Grazing Incidence Spectrometer (GIS) and the Normal Incidence Spectrometer (NIS). The emission lines considered in the present work are all from the NIS: the magnesium lines all lie in the NIS1 waveband (308--379~\AA) and the neon and oxygen lines all lie in the NIS2 waveband (513--633~\AA). A complete list of the atomic transitions is given in Table~\ref{tbl.lines}. Due to self-blending at the CDS spectral resolution, individual emission features correspond to multiple atomic transitions for some ions. In these cases the emissivities for the individual transitions are summed in the atomic models.

\begin{deluxetable}{llc}
\tablecaption{Atomic transitions used for the analysis.\label{tbl.lines}}
\tablehead{Ion & Transition  & $\lambda$/\AA }
\startdata
\ion{O}{iii} & $2s^22p^2$ $^1S_0$ -- $2s2p^3$ $^1P_1$ & 599.59\\
\ion{O}{iv} & $2s^22p$ $^2P_{1/2}$ -- $2s2p^2$ $^2P_{3/2}$ & 553.33 \\
 & $2s^22p$ $^2P_{1/2}$ -- $2s2p^2$ $^2P_{1/2}$ & 554.08 \\
 & $2s^22p$ $^2P_{3/2}$ -- $2s2p^2$ $^2P_{3/2}$ & 554.51 \\
 & $2s^22p$ $^2P_{3/2}$ -- $2s2p^2$ $^2P_{1/2}$ & 555.26 \\
\ion{O}{v} & $2s^2$ $^1S_0$ -- $2s2p$ $^1P_1$ & 629.73 \\
\ion{Ne}{iv} & $2s^22p^3$ $^4S_{3/2}$ -- $2s2p^4$ $^4P_{5/2}$ & 543.89 \\
\ion{Ne}{v} & $2s^22p^2$ $^3P_2$ -- $2s2p^3$ $^3D_1$ & 572.03\\
\ion{Ne}{vi} & $2s^22p$ $^2P_{3/2}$ -- $2s2p^2$ $^2D_{3/2}$ & 562.71 \\
 & $2s^22p$ $^2P_{3/2}$ -- $2s2p^2$ $^2D_{5/2}$ & 562.81 \\
\ion{Ne}{vii} & $2s2p$ $^3P_1$ -- $2p^2$ $^3P_1$ & 561.38\\
 & $2s2p$ $^3P_2$ -- $2p^2$ $^3P_2$ & 561.73\\
\ion{Mg}{v} & $2s^22p^4$ $^3P_2$ -- $2s2p^5$ $^3P_2$ & 353.09 \\
 & $2s^22p^4$ $^3P_1$ -- $2s2p^5$ $^3P_1$ & 353.30 \\
\ion{Mg}{vi} & $2s^22p^3$ $^2D_{5/2}$ -- $2s2p^4$ $^2D_{3/2}$ & 349.11 \\
 & $2s^22p^3$ $^2D_{3/2}$ -- $2s2p^4$ $^2D_{3/2}$ & 349.13 \\
 & $2s^22p^3$ $^2D_{5/2}$ -- $2s2p^4$ $^2D_{5/2}$ & 349.16 \\
 & $2s^22p^3$ $^2D_{3/2}$ -- $2s2p^4$ $^2D_{5/2}$ & 349.18 \\
\ion{Mg}{vii} & $2s^22p^2$ $^3P_2$ -- $2s2p^3$ $^3P_2$ & 367.68\\
 & $2s^22p^2$ $^3P_2$ -- $2s2p^3$ $^3P_1$ & 367.69\\
\ion{Mg}{viii} & $2s^22p$ $^2P_{3/2}$ -- $2s2p^2$ $^2P_{3/2}$ & 315.02 \\
\enddata
\end{deluxetable}

The  emission line intensity measurements of \citet{young05-fip,2005A&A...444L..45Y} are used, and no new measurements were performed. The quiet Sun was divided into equal spatial areas of network and supergranule cell centers based on the intensity of  \ion{O}{v} \lam629.7  (formed at 0.2~MK) and intensities were measured for 24 datasets between 1996 March and 1998 June. This period was prior to the temporary loss of SOHO in 1998\footnote{http://www.esa.int/esapub/bulletin/bullet97/vandenbu.pdf}, after which the NIS line profiles were significantly distorted, making measurements of weak lines difficult. For the present work we derive abundance ratios for the average quiet Sun in addition to network and cell centers, and these were derived simply by averaging the network and cell center intensities, since the two regions have equal spatial areas. We refer the reader to \citet{young05-fip,2005A&A...444L..45Y} for further details on how the data were processed and how the lines were measured. 

% For this work we provide average quiet Sun abundance ratios that were obtained simply by averaging the network and cell center We refer the reader to \citet{young05-fip,2005A&A...444L..45Y} for details on how the data were processed and how the lines were measured. We note that only the observation period 1996 March to 1998 June was considered as the NIS line profiles were significantly distorted after the temporary loss of SOHO in 1998.

The standard radiometric calibration of NIS has not changed since 2002, and so the intensities measured by \citet{young05-fip,2005A&A...444L..45Y} are still valid. An alternative calibration was presented by \citet{2010A&A...518A..49D} and we find differences of 15\%\ or less with the standard calibration for the time period considered here, which is within the uncertainty of the NIS calibration.

\section{Photospheric abundances}\label{sect.abun}

In this section the current status of the photospheric abundances of oxygen, neon and magnesium is  discussed. The notation used for  abundances is as follows: the abundance, $\epsilon$, of an element X is defined as $N({\rm X})/N({\rm H})$ where $N$ is the number density. The abundance on a logarithmic scale is $A({\rm X})=12+\log\,\epsilon({\rm X})$, and it is this value that is typically given in compilations. 

% For the Mg/Ne abundance ratio we are interested in whether it takes an enhanced value in the quiet Sun compared with the photospheric ratio. However, the neon photospheric abundance can not be directly measured and so has to be inferred through other means. There is some variation in how this is done in the abundance compilations in the literature, and we summarize the two most recent compilations below.

% There remains some controversy over solar photospheric abundances, resulting in different compilations that are used in solar physics and astronomy. Here we focus on the values for O, Ne and Mg. The compilation of 

\citet{asplund09} presented abundances mostly derived from results of their 3D magnetoconvection simulations of photospheric absorption lines. Their recommended value for oxygen was $A({\rm O})=8.69\pm 0.05$.  Updates for elements heavier than oxygen were published by \citet{2015A&A...573A..25S}, \citet{2015A&A...573A..26S} and \citet{2015A&A...573A..27G}, and the magnesium abundance was $7.59\pm 0.04$. The neon abundance remained unchanged compared to \citet{asplund09}, and the value of $7.93\pm 0.10$ was derived from the oxygen abundance by applying the Ne/O ratio of \citet{2005A&A...444L..45Y}. With these values the photospheric Mg/Ne ratio is 0.47. We refer to the updated \citet{asplund09} compilation as "Asp-Sco" in the following text.

A distinct abundance compilation was created by \citet{lodders09} who were more cautious in the use of abundances derived from 3D models. The oxygen abundance was taken as $8.73\pm 0.07$, a little higher than \citet{asplund09} but consistent within the error bars. For magnesium, \citet{lodders09} used the value of $7.54\pm 0.06$ from \citet{1979LIACo..22..117H}---see also \citet{2001AIPC..598...23H}---which is close to the meteoritic value of 7.53. The most significant difference compared to \citet{asplund09} is the neon abundance, with a value of $8.05\pm 0.10$. This is an average of the value of \citet{2008A&A...487..307M}, derived from a sample of B stars, and the solar flare value of \citet{2007ApJ...659..743L}. Both of these are absolute abundance measurements, i.e., relative to hydrogen.

\citet{caffau11} calculated photospheric abundances for 12 elements using the CO5BOLD code, and suggested that these replace the values in the \citet{lodders09} compilation. Mg and Ne were not updated, but oxygen was increased to $8.76\pm 0.07$. We refer to the hybrid Lodders--Caffau compilation as "Lod-Caf" in the following text. 

We summarize the different abundance values and their corresponding ratios in Table~\ref{tbl.phot}, where we also compare with the compilation of \citet{grevesse98} that was made prior to the use of 3D models for photospheric line studies. 

\begin{deluxetable*}{lccccc}
\tablecaption{Abundances and abundance ratios from select compilations.\label{tbl.phot}}
\tablehead{Compilation & $A$(O) & $A$(Ne) & $A$(Mg) & $R_{\rm Mg/Ne}$ & $R_{\rm Ne/O}$}
\startdata
\citet{grevesse98} & $8.83\pm 0.06$ & $8.08\pm 0.06$ & $7.58\pm 0.05$ & $0.316\pm 0.061$ & $0.178\pm 0.037$ \\
\citet{asplund09} & $8.69\pm 0.05$ & $7.93\pm 0.10$ & $7.60\pm 0.04$ & $0.468\pm 0.129$ & $0.174\pm 0.050$\\
Asp-Sco & $8.69\pm 0.05$ & $7.93\pm 0.10$ & $7.59\pm 0.04$ & $0.457\pm 0.126$ & $0.174\pm 0.050$ \\
\citet{lodders09} & $8.73\pm 0.07$ & $8.05\pm 0.10$ & $7.54\pm 0.06$ & $0.309\pm 0.092$ & $0.209\pm 0.065$ \\
Lod-Caf & $8.76\pm 0.07$ & $8.05\pm 0.10$ & $7.54\pm 0.06$ & $0.309\pm 0.092$ & $0.195\pm 0.061$ \\
\enddata
\end{deluxetable*}

% The increase of the photospheric Mg/Ne ratio in the \citet{asplund09} compilation resulted from the neon abundance being tied to oxygen, and so the decrease of the oxygen abundance by 0.14~dex resulted in a similar decrease of neon. By contrast, the magnesium abundance remained almost the same compared to \citet{grevesse98}.

% For the \citet{lodders09} compilation, the neon abundance was not tied to oxygen, instead the average of the absolute measurements of \citet{2008A&A...487..307M} and \citet{2007ApJ...659..743L} were used. After a small increase in the oxygen abundance from \citet{caffau11}, the Mg/Ne abundance became close to the \citet{grevesse98} value again. 

% The wide range of photospheric abundances in the literature is not of direct importance to the present work as we seek, firstly, the transition region Ne/O ratio which we assume is a measure of the photospheric ratio; and, secondly, the Mg/Ne ratio to determine if the effect is present in the quiet Sun. 

\section{The standard solar model and helioseismology results}\label{sect.helio}

% The photospheric abundances from the compilations discussed in the previous sections are derived mostly from the modeling of photospheric absorption lines, which requires models of radiative transfer and convection in the outer layer of the Sun. 

Prior to the updates to the photospheric abundances of the key elements carbon, nitrogen and oxygen in the early 2000's, the predictions of the standard solar model (SSM) of the solar interior were in excellent agreement with measurements of solar parameters from helioseismology. The updated abundances led to discrepancies that have not yet been resolved, and some authors argue that the discrepancies demonstrate that the updated abundances are incorrect. For example, \citet{2014ApJ...787...13V} treat the ratio of volatiles to refractory elements as a free parameter in reconciling the SSM with the helioseismology and solar neutrino results and find the ratio is more consistent with the \citet{grevesse98} compilation than the \citet{asplund09} compilation. Until this situation is resolved there will remain a cloud of uncertainty over the photospheric abundances of key elements such as oxygen and neon,  and these cause problems when trying to measure the FIP bias in solar atmosphere plasma. 

The fact that the photospheric neon abundance is only measured indirectly led \citet{2005ApJ...620L.129A} and \citet{2005ApJ...631.1281B} to suggest that the modeling--observation discrepancy could be resolved if the neon abundance was actually two to three times higher than previously thought. A study of stellar coronal abundances measured from \emph{Chandra} spectra \citep{2005Natur.436..525D} suggested this was plausible, but the subsequent study of \citet{2008A&A...486..995R} found that the Ne/O ratio was correlated with stellar activity, with low activity stars being consistent with the accepted solar Ne/O ratio. There remains the key question of whether the coronal Ne/O abundance actually represents the photospheric value, and we will return to this issue in Sect.~\ref{sect.cor}.

% Given the interest in neon for the present work, we note that \citet{2014ApJ...787...13V} considered whether treating the Ne/O as a free parameter could enable modeling--observation problem to be resolved while retaining the 

% The standard solar model of the Sun's interior makes use of these abundances for computing opacity 

\section{Ionization fractions}\label{sect.ionfrac}

A standard assumption in astrophysics is that the ionization fractions of an element at any given temperature are independent of the electron density and they are often referred to as "zero-density ion fractions" as they are correct in the limiting case of density tending to zero.
Prior to CHIANTI 6, scientists typically took the zero-density ion fractions from standard tabulations such as \citet{1985A&AS...60..425A}, \citet{mazzotta98} and \citet{2009ApJ...691.1540B}. With CHIANTI 6 \citep{chianti6}, critically-assessed ionization and recombination rates were added to the database and used to compute a new ionization equilibrium dataset that is now the default option for CHIANTI. The analysis of \citet{young05-fip,2005A&A...444L..45Y} took place prior to this change, and the \citet{mazzotta98} dataset was used. 

Density effects can modify the equilibrium ion fractions through  the suppression of dielectronic recombination (DR) rates, which is caused by electron collisions de-populating highly-excited states prior to recombination. \citet{nikolic13} provided formulae for computing the suppression factors for astrophysically important ions, and these formulae have been implemented by the author to compute density and pressure sensitive ion fraction tables derived by applying the suppression factors to the DR rates in CHIANTI 8 \citep{chianti8}. One feature of the \citet{nikolic13} suppression factor formulae is an unphysical  discontinuity of a factor two for the isoelectronic sequences hydrogen through boron \citep[see Sect.~2.2.1 of][]{nikolic13}, and for \ion{O}{iv} and \ion{O}{v} the discontinuities occur close to the temperatures of maximum ionization  of the ions.
% We note that one feature of the \citet{nikolic13} formulae was that the suppression factors were discontinuous for the isoelectronic sequences hydrogen through boron, which resulted from a factor two being applied below a certain temperature. 
D.~Nikoli\'c (private communication, 2017) provided the author with an updated formula that smooths out this discontinuity and also provides an improved fit to the original data for these sequences. This formula was applied in the present work.

\begin{figure}[h]
\epsscale{1.0}
\plotone{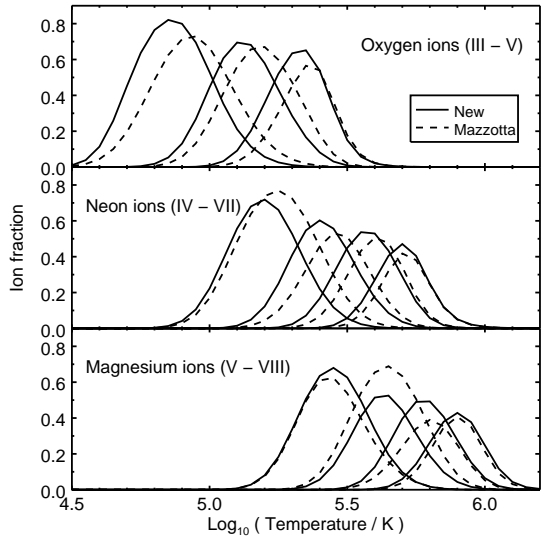}
\caption{Ion fraction plots for the oxygen, neon and magnesium ions studied in the present work. The solid lines show the curves derived for a pressure of $10^{14.5}$~K~cm$^{-3}$; the dashed lines show the curves from \citet{mazzotta98}.}
\label{fig.fracs}
\end{figure}

Figure~\ref{fig.fracs} compares the ion fractions computed at a pressure of $10^{14.5}$~K~cm$^{-3}$ with those from \citet{mazzotta98}, and significant differences are found for several of the ions. For example, \ion{O}{iii} and \ion{O}{iv} are shifted to lower temperatures by 0.1~dex, and the ion fraction of \ion{Mg}{vi} is 30\%\ lower. The differences are a combination of  the updated ionization and recombination data distributed with CHIANTI and the changes introduced by DR suppression. Appendix~\ref{sect.app} discusses the differences between the new pressure equilibrium dataset and the CHIANTI 8 file.

\section{Other atomic data}\label{sect.atom}

The atomic data for computing emissivities are all taken from CHIANTI 8 \citep{chianti8}, with the exception of \ion{Ne}{vi}. An error in the data files was found for this ion, whereby levels 6 and 7 ($2s2p^2$ $^2D_{3/2,5/2}$) were mis-labeled, and these levels give rise to the \lam562.71 and \lam562.81 transitions that are measured in the CDS spectra. The modified \ion{Ne}{vi} files will be released with the next version of the database. 

The initial study of the Mg/Ne ratio \citep{young05-fip} made use of CHIANTI 4, and the results were updated by \citet{2006ESASP.617E..47Y} using CHIANTI 5. This version was also used by \citet{2005A&A...444L..45Y} for the Ne/O ratio. Since CHIANTI 5, all of the ions considered in the present work have been updated, except for \ion{Ne}{vii}, \ion{Mg}{v} and \ion{Mg}{vi}. Further details are given in the papers that describe versions 6, 7, 7.1 and 8 of the database \citep[][respectively]{chianti6,chianti7,chianti71,chianti8}.

\section{Method}\label{sect.method}

The model for the emission line intensities is a slightly modified version of the one used by \citet{young05-fip, 2005A&A...444L..45Y}, which writes the intensity, $I$,  as a sum of isothermal plasma components:
\begin{equation}\label{eq.int}
  I = \epsilon \sum_k G(T_k,N_{{\rm e},k}) N_{{\rm e},k} N_{{\rm H},k} h_k
\end{equation}
where $G$ is the contribution function, $N_{\rm H}$ and $N_{\rm e}$ are the number densities of hydrogen and electrons, respectively, $T$ is the plasma temperature, and $h$ is the plasma column depth. The sum is performed over temperatures uniformly spaced on a $\log\,T$ grid. The contribution function contains the atomic physics parameters, which are all taken from CHIANTI 8, and it is derived using the GOFNT procedure in the CHIANTI IDL software with the /NOABUND keyword set and with the ion fraction file described in Sect.~\ref{sect.ionfrac}.

A constant pressure, $P=N_{\rm e}T=10^{14.5}$~K~cm$^{-3}$, is assumed, based on the coronal density measured in the quiet Sun by \citet{young05-fip}. This was derived using a \ion{Si}{ix} density diagnostic and only small changes to the CHIANTI atomic model have been since CHIANTI 4 \citep[see][]{chianti6,chianti7} and these had a negligible impact on the diagnostic. For the present work the temperature grid has a spacing of 0.05~dex.
 The previous work used 0.1~dex intervals, and the change reflects the improvement in resolution of the ionization equilibrium files distributed with the more recent versions of CHIANTI.

\begin{figure}[h]
\plotone{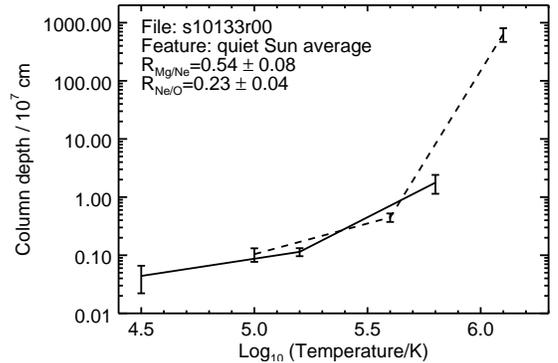}
\caption{Column depth distributions for one CDS dataset are shown, as derived with the Ne/O (solid line) and Mg/Ne (dashed line) methods. The points with the error bars correspond to the $p_1$, $p_2$ and $p_3$ values that are derived from the minimization method---see main text.}
\label{fig.coldepth}
\end{figure}

In order that a $\chi^2$ minimization technique can be applied, we define the column depth distribution simply by making the $h$ values at the start, middle and end of the temperature range to be free parameters. We refer to these three values as $p_1$, $p_2$ and $p_3$ and the temperatures at which they are defined are $\log\,T=5.0$, 5.6 and 6.1 for the Mg/Ne method, and 4.5, 5.2 and 5.8 for the Ne/O method.  All other $h$ values on the temperature grid are obtained by drawing straight lines between $p_1$ and $p_2$, and $p_2$ and $p_3$ in $\log\,T$--$\log\,h$ space. Figure~\ref{fig.coldepth} shows examples of the derived $h$ distributions. For the Mg/Ne calculation, the four fitting parameters are adjusted to achieve the best agreement with the eight observed Mg and Ne emission lines, while for the Ne/O calculation there are six emission lines (\ion{Ne}{vii} is not included as it has negligible overlap with the three oxygen ions---see Figure~\ref{fig.fracs}). There is one emission line from each ion, and the  atomic transitions  are given in Table~\ref{tbl.lines}, where we note that some lines consist of multiple transitions at the CDS spectral resolution. The minimization procedure was performed with the MPFIT IDL routines \citep{2009ASPC..411..251M}, which yielded $1\sigma$ error bars on the abundance ratios and $p_i$ values.

If we consider the example of determining the Ne/O relative abundance ratio, then we write the intensities of the neon and oxygen lines at wavelengths $\lambda$ by 
\begin{eqnarray}
I_{\lambda,{\rm Ne}} & = & R_{\rm Ne/O} \epsilon({\rm O}) f_\lambda(p_1,p_2,p_3) \label{eq.ne}\\
I_{\lambda,{\rm O}} & = &  \epsilon({\rm O}) f_\lambda(p_1,p_2,p_3) \label{eq.o}
\end{eqnarray}
% \begin{equation}\label{eq.ne}
% I_{\lambda,{\rm Ne}} = R_{\rm Ne/O} \epsilon({\rm O}) f_\lambda(p_1,p_2,p_3)
% \end{equation}
% \begin{equation}\label{eq.o}
% I_{\lambda,{\rm O}} =  \epsilon({\rm O}) f_\lambda(p_1,p_2,p_3)
% \end{equation}
%
where we use $f_\lambda$ to indicate the sum in Eq.~\ref{eq.int}, and $R_{\rm Ne/O}=\epsilon({\rm Ne})/\epsilon({\rm O})$. The free parameters are then $R_{\rm Ne/O}$, $p_1$, $p_2$ and $p_3$, and a least squares fit is performed to minimize the difference between the observed intensities and the $I_\lambda$ values. Examples of the $p_1$, $p_2$ and $p_3$ values derived for the Mg/Ne and Ne/O methods for one CDS dataset are shown in Figure~\ref{fig.coldepth}.

We highlight that the abundance of oxygen enters into Eqs.~\ref{eq.ne} and \ref{eq.o} as a fixed parameter. We take this, and the abundance of neon for the Mg/Ne case, from the Lod-Caf abundance compilation. This choice does not affect the final abundance ratio, as can be seen by considering Eq.~\ref{eq.o}. Scaling $\epsilon$(O) by some factor $\alpha$ requires that $f(\lambda)$ be scaled downwards by $\alpha$, and thus $R_{\rm Ne/O}$ in Eq.~\ref{eq.ne} is unaffected. The derived column depths are affected, however.
% but it does affect the derived column depths. 
For example, if the Asp-Sco neon abundance was used instead for the Mg/Ne method, then the column depths would be 32\%\ larger to compensate for the lower neon abundance (Table~\ref{tbl.phot}).

% The only free parameters for the column depth are the values at the ends of the temperature range, and a mid temperature point. The temperatures are given in Table~\ref{tbl.params} for the Mg/Ne and Ne/O methods. The remaining $h_k$ values are derived by drawing straight lines in $\log\,T$--$\log\,h$ space between the two pairs of points 

% If we consider the specific case of the intensities of lines of wavelength, $\lambda$, belonging to neon and oxygen ions, then
% %
% \begin{equation}
% I_{\lambda,{\rm Ne}} = R_{\rm Ne/O} A(O) f_\lambda(p_1,p_2,p_3)
% \end{equation}
% \begin{equation}
% I_{\lambda,{\rm O}} =  A(O) f_\lambda(p_1,p_2,p_3)
% \end{equation}
% %
% then we apply a minimization method to best reproduce the observed intensities by varying $R_{\rm Ne/O}$, $p_1$, $p_2$ and $p_3$.

% \begin{deluxetable}{llll}
% \tablecaption{Temperatures at which the $p_1$, $p_2$ and $p_3$ values are defined.\label{tbl.params}}
% \tablehead{&\multicolumn{3}{c}{$\log\,(T/{\rm K})$} \\ 
% \cline{2-4}
% Method & $p_1$ & $p_2$ & $p_3$}
% \startdata
% Mg/Ne  & 5.0 & 5.6 & 6.1 \\
% Ne/O   & 4.5 & 5.2 & 5.8 \\
% \enddata
% \end{deluxetable}

One change compared to the earlier analyses is the use of the correct temperature-dependent
value for the ratio $N_{\rm H}$/$N_{\rm e}$ that is derived using the
element abundance and ionization equilibrium files using the CHIANTI
routine PROTON\_DENS. The ratio shows a negligible variation between the various photospheric abundance datasets, and we choose the Lod-Caf  set here. Finally, a
mistake in the previous method was corrected whereby  line emissivities
were computed at constant density rather than constant pressure,
although this has a minor effect on the results.

% A $\chi^2$ minimization procedure is applied such that the three column depths and the relative abundance are varied in order to best reproduce the intensities of the observed lines, for the Ne/O and Mg/Ne ratios separately. Examples of the derived column depth distributions are shown in Figure~\ref{fig.coldepth}, which should be compared with Figure~3 of \citet{2005A&A...444L..45Y}.

\section{Results}\label{sect.results}

For each of the 24 CDS datasets we apply the minimization method to determine the Ne/O and Mg/Ne relative abundance ratios for the average quiet Sun, and the network and cell center regions separately, using the sets of oxygen, neon and magnesium lines. \citet{young05-fip,2005A&A...444L..45Y} demonstrated that the ratios show no systematic variation with time and this does not change here, so we only discuss the ratio values averaged over the 24 datasets.

\subsection{The Ne/O ratio}

The average Ne/O abundance ratios for network and cell center are $0.239\pm 0.010$ and $0.258\pm 0.012$, respectively. The difference between these values only has a 1.2~$\sigma$ significance and so for the remainder of this section we consider only the average quiet Sun value, which is $0.244\pm 0.010$. This is about 40\%\ higher than that derived by \citet{2005A&A...444L..45Y}, and the difference mostly lies with the change in the ion fractions. If we use the original \citet{mazzotta98} ion fraction file, but interpolated onto a 0.05~dex grid in temperature (to be compatible with the modern files), then we find a ratio of 0.183, which is only 10\% higher than the \citet{2005A&A...444L..45Y} value. With the zero-density ion fraction file distributed with CHIANTI 8, the value is 0.237 and so the largest part of the increase comes from the changes to ionization and recombination rates between \citet{mazzotta98} and the CHIANTI 8 ion balance calculation.

% \citet{2005A&A...444L..45Y} estimated the uncertainties in the ionization fractions by shifting the neon ion fractions up and down in temperature by 0.1~dex, and repeating this here gives ratios of 0.242 and 0.257 for an upwards shift, and 0.274 and 0.304 for a downward shift, values within the uncertainties of the unshifted results.

The small uncertainty on the Ne/O ratio comes from the standard deviation of the values derived from the 24 datasets. The uncertainties in the atomic data are likely the biggest source of errors, as reflected in the large change to the Ne/O ratio found here, but they are difficult to quantify.  Here we assign a 20\%\ uncertainty that, when combined with the statistical error, gives a final quiet Sun Ne/O ratio of $0.244\pm 0.050$.

If we now interpret this average quiet Sun value as representing the photospheric ratio, then we have the choice of the \citet{asplund09} oxygen value, or the \citet{caffau11} value. The former results in a neon abundance of $A({\rm Ne})=8.08\pm 0.09$, while the latter yields $A({\rm Ne})=8.15\pm 0.10$.

Noting that \citet{asplund09} used our previous result to derive the neon abundance, we suggest this compilation be updated with the value $A({\rm Ne})=8.08\pm 0.09$.
As mentioned in Sect.~\ref{sect.abun}, the compilation of \citet{lodders09} took the neon abundance from two absolute measurements, rather than ratio values.

% The significantly larger Ne/O ratio found here has consequences for the photospheric neon abundance if we recall that \citet{asplund09} used the result from \citet{2005A&A...444L..45Y} to derive the photospheric abundance of neon from the abundance of oxygen. If we use the present Ne/O ratio, then we find $A(Ne)=8.13\pm 0.10$ using the \citet{asplund09} oxygen abundance. 

% In arriving at final values and uncertainties, the procedure is as follows. The effects of DR suppression on the ion fractions may be over-estimated as no account of the density effects of ionization are included (specifically effects due to ionization out of metastables). Therefore I choose to take the average of the 

% \begin{figure}[h]
% \epsscale{1.0}
% \plotone{plot_ne_o_ratio.eps}
% \caption{Ne/O relative abundance ratios derived from CDS data for quiet Sun network (top) and cell center regions (bottom) as a function of time. The solid and dashed horizontal lines show the photospheric ratios from the Asp-Sco  and Lod-Caf  compilations, respectively.}
% \label{fig.neo}
% \end{figure}

\subsection{The Mg/Ne ratio}

The average Mg/Ne ratios for quiet Sun network and cell center regions are $0.497\pm 0.035$ and $0.609\pm 0.077$, which are  15--25\%\ larger than those of \citet{young05-fip}, but only marginally different to the updated values of \citet{2006ESASP.617E..47Y}. Thus the changes in the ionization fractions since \citet{mazzotta98} shown in Figure~\ref{fig.fracs} do not have a significant impact on the Mg/Ne  ratio. The difference between the network and cell center values is only at the 1.3~$\sigma$ level, and if we use the average quiet Sun intensities we find a ratio of $0.519\pm 0.041$. Applying a 20\%\ uncertainty on the atomic data gives a final Mg/Ne ratio of $0.519\pm 0.112$.

The Mg/Ne FIP bias in the quiet Sun depends on the choice of photospheric abundances. For the Asp-Sco abundance compilation, we update the neon abundance as described in the previous section, which gives a photospheric Mg/Ne ratio of $0.326\pm 0.084$. The quiet Sun FIP bias is then 
 $1.59\pm 0.54$, and for the Lod-Caf abundance set it is $1.68\pm 0.62$. For simplicity we can state the FIP bias as $1.6\pm 0.6$, independently of which abundance set is used. We thus find  
a small FIP bias in the upper transition region of the quiet Sun, although the ratio values are only 1~$\sigma$ above the photospheric ratios.

% \begin{figure}[h]
% \epsscale{1.0}
% \plotone{plot_mg_ne_ratio.eps}
% \caption{Mg/Ne relative abundance ratios derived from CDS data for quiet Sun network (top) and cell center regions (bottom) as a function of time. The solid and dashed horizontal lines show the photospheric ratios from the Asp-Sco  and Lod-Caf  compilations, respectively.}
% \label{fig.mgne}
% \end{figure}

\section{Comparison with coronal Ne/O results}\label{sect.cor}

In this section we discuss recent progress in measuring the coronal Ne/O ratio. We first
% In comparing the coronal neon abundances with those derived here, we 
highlight that transition region emission is generally considered to come from dynamic structures that \emph{are not physically connected to the corona} \citep{1983ApJ...275..367F}.
Coronal structures pass through transition region temperatures as they connect to the photosphere, but this emission is much weaker than the disconnected structures. Direct evidence for these disconnected structures was recently presented from high spatial resolution UV observations obtained by the IRIS satellite \citep{2014Sci...346E.315H}, which also showed that they are highly dynamic with time-scales of minutes. If the physical mechanism that causes the FIP effect is assumed to operate over a period of time, then it is more likely that the dynamic transition region structures will show a smaller FIP bias than the longer-lived coronal structures, as recently suggested by \citet{2016ApJ...824...56W}.

There have been several papers that have derived neon abundances from solar and stellar coronal spectra. \citet{2007ApJ...659..743L} and \citet{2015ApJ...800..110L} used UV spectra from the Solar Ultraviolet Measurements of Emitted Radiation (SUMER) experiment on board SOHO to derive absolute neon abundances in off-limb coronal regions. The former work used a \ion{Ne}{ix} line formed at 4~MK and produced during the decay phase of a solar flare. By comparing with the strength of the free-free continuum, a value of $A({\rm Ne})=8.11\pm 0.12$ was derived and this was used in the \citet{lodders09} abundance compilation.

A different approach was adopted by \citet{2015ApJ...800..110L} who used lines of \ion{Ne}{viii} and \ion{O}{vi} to measure the Ne/O ratio, and also used lines of \ion{H}{i} to yield absolute abundances for Ne and O. The \ion{Ne}{viii} and \ion{O}{vi} lines were measured in the off-limb corona where they are formed at about 1.3~MK, significantly higher than their $T_{\rm max}$ values of 0.63 and 0.28~MK, respectively, which is enabled by the extended high temperature tails of lithium-like ions. Abundances were derived from many datasets over the period 1996 to 2008, and time variations in the Ne/O ratio were found. Although the ratio was relatively constant during 1996--2005 (solar cycle 23), it rose significantly as the Sun entered the minimum of solar cycle 24. Interestingly, \citet{2015ApJ...800..110L} argue that the high value of $R_{\rm Ne/O}=0.25\pm 0.05$---a value consistent with that derived here---found at the end of the sequence in 2008 is the best representation of the true solar photospheric value due to the very low solar activity at this time, which may have depressed the FIP bias mechanism.

\citet{2011ApJ...743...22D} reassessed the datasets studied by \citet{1992ApJ...389..764M} and \citet{2005ApJ...634L.197S}, processing \ion{Ne}{ix} and \ion{O}{viii} X-ray line intensities to derive Ne/O ratios. Consistent with the earlier authors, he found evidence that the Ne/O ratio varies from dataset to dataset with an average close to the values from the \citet{grevesse98} and \citet{asplund09} datasets (and so lower than the value found here). A new result was the finding of a correlation with active region temperature. This was interpreted as supporting a result from stellar X-ray studies that the Ne/O ratio increases with stellar activity \citep{2008A&A...486..995R}. We note however, that this contradicts the result of \citet{2015ApJ...800..110L} who suggested the Ne/O ratio is higher at times of low solar activity. 

The conclusion that the coronal Ne/O varies from region to region and is possibly related to activity suggests that it is not a reliable measure of the photospheric ratio. The use of transition region data from quiet Sun in the present analysis is suggested here to be preferable, especially taking into account the low FIP bias found from Mg/Ne from the same datasets.

% Stellar coronal spectra also provide a means for measuring the neon abundance as X-ray spectra such as obtained by Chandra and XMM-Newton give access to strong lines of \ion{Ne}{ix} and \ion{Ne}{x}. The first survey of \citet{2005Natur.436..525D} gave a value of $R_{\rm Ne/O}=0.40$, significantly higher than typical solar values. 

% In comparing the coronal neon abundances with those derived here, we highlight that transition region emission is generally considered to come from dynamic structures that \emph{are not physically connected to the corona}. The footpoints of coronal structures pass through transition region temperatures as they connect to the photosphere, but this emission is much weaker than the disconnected structures. Evidence for these disconnected structures was recently presented from high spatial resolution UV observations obtained by the IRIS satellite \citep{2014Sci...346E.315H}. As plasma structures around $10^5$~K are naturally unstable, then they are short-lived and thus unlikely to show a FIP effect. 

\section{Summary}\label{sect.summary}

\begin{deluxetable}{lcc}
\tablecaption{Summary of abundance results.\label{tbl.results}}
\tablehead{Quantity & Value} 
\startdata
$R_{\rm Mg/Ne}$ & $0.519\pm 0.112$ \\
$R_{\rm Ne/O}$ & $0.244 \pm 0.050$ \\
FIP bias (Mg/Ne)   & $1.68\pm 0.62$ &(Lod-Caf) \\
                 & $1.59\pm 0.54$ &(Asp-Sco) \\
$A$(Ne)  & $8.15\pm 0.10$  &(Lod-Caf) \\
                & $8.08\pm 0.09$  & (Asp-Sco)\\
\enddata
\end{deluxetable}

The results from this work are summarized in Table~\ref{tbl.results}. The Ne/O and Mg/Ne relative abundance ratios in the upper transition region of the quiet Sun have been re-evaluated by applying updated atomic data to previously published measurements. The key result is that the Ne/O ratio is 40\%\ higher than found previously, a change driven by updates to the equilibrium ionization fractions of the neon and oxygen ions.

% \begin{deluxetable}{ccc}
% \tablecaption{Summary of abundance results.\label{tbl.results}}
% \tablehead{&Lod-Caf & Asp-Sco} 
% \startdata
% Mg/Ne FIP bias   & $1.68\pm 0.62$ & $1.59\pm 0.54$ \\
% $\epsilon$(Ne)  & $8.15\pm 0.10$ & $8.08\pm 0.09$ \\
% \enddata
% \end{deluxetable}

If we assume the quiet Sun Ne/O ratio represents the photospheric ratio, then the photospheric abundance of oxygen from \citet{asplund09} implies a neon abundance of $A({\rm Ne})=8.08\pm 0.09$. If we use the higher oxygen abundance  derived by \citet{caffau11}, then we have $A({\rm Ne})=8.15\pm 0.10$. 

The Mg/Ne quiet Sun ratios are very similar to those derived by \citet{2006ESASP.617E..47Y}, despite the updates in the ionization fractions. If 
we use the updated photospheric neon abundances found here, then a larger  Mg/Ne FIP bias of $1.6\pm 0.6$ is found compared to the earlier work. This remains significantly smaller than the canonical FIP bias values of 4 and 3 used in the reviews of \citet{feldman92} and \citet{schmelz12}, respectively. The abundance results are thus consistent with the notion that the observed transition region is predominantly disconnected from the corona.

% However, the change in the photospheric neon abundance suggested by the Ne/O ratio is significant for determining the Mg/Ne FIP bias. In contrast to \citet{2006ESASP.617E..47Y} we now find small FIP biases of 1.7 and 2.1 for network and cell center regions. These values are however smaller than the canonical FIP bias value of 3.0 that was adopted by \citet{schmelz12}, suggesting the FIP effect is less effective in quiet Sun transition region plasma.

% The Mg/Ne ratios in quiet Sun network and cell center regions derived here are very similar to those derived by \citet{2006ESASP.617E..47Y}, despite the updates in the ionization fractions. However, the change in the photospheric neon abundance suggested by the Ne/O ratio is significant for determining the Mg/Ne FIP bias. In contrast to \citet{2006ESASP.617E..47Y} we now find small FIP biases of 1.7 and 2.1 for network and cell center regions. These values are however smaller than the canonical FIP bias value of 3.0 that was adopted by \citet{schmelz12}, suggesting the FIP effect is less effective in quiet Sun transition region plasma.

\acknowledgments
The author acknowledges funding from NASA grants NNX15AF48G and NNX15AF25G. SOHO is a project of international cooperation between ESA and NASA. CHIANTI is a collaborative project involving George Mason University, the University of Michigan (USA) and the University of Cambridge (UK). The referee is thanked for valuable comments. 

\facilities{SOHO(CDS)} 
\software{Solarsoft, CHIANTI}

\bibliographystyle{aasjournal}
\bibliography{main}

\appendix

\section{Effects of DR suppression on the CHIANTI ion balance file}\label{sect.app}

Figure~\ref{fig.fracs} showed the differences between the equilibrium ion fractions adopted in the present work and the ion fractions from \citet{mazzotta98} that were used by \citet{young05-fip,2005A&A...444L..45Y}. In this appendix we highlight the effect caused by  DR suppression on the current CHIANTI 8 zero-density equilibrium ion fractions, and in particular how it affects certain ions. 

\begin{figure}[h]
\epsscale{0.5}
\plotone{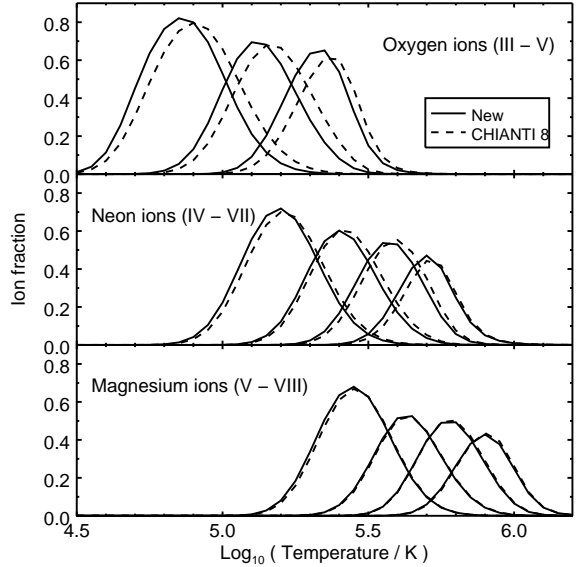}
\caption{Ion fraction plots for the oxygen, neon and magnesium ions studied in the present work. The solid lines show the curves derived for a pressure of $10^{14.5}$~K~cm$^{-3}$; the dashed lines show the curves from distributed with CHIANTI 8.}
\label{fig.fracs2}
\end{figure}

Figure~\ref{fig.fracs2} is an analogous plot to Figure~\ref{fig.fracs}, but comparing the ion fractions at $P=10^{14.5}$~K~cm$^{-3}$ with the zero-density CHIANTI 8 file from which it is derived. Since the ionization rates are unaffected, the suppression of the DR rates results in the ion fraction curves moving to lower temperatures. This is most strongly seen for the oxygen ions, whereas the magnesium ions are relatively unaffected. The peaks of the ion fraction curves  show changes of at most a few percent.

Table~\ref{tbl.act} gives a list of the ions used in the present work, with their temperatures of maximum ionization ($T_{\rm max}$), and the factors, $S$, by which DR is suppressed at the ions' $T_{\rm max}$ values. The formulae of \citet{nikolic13} only correct an ion's DR rate above a certain density that is referred to as the activation density, and these values  are also given in Table~\ref{tbl.act}. We note that the  activation density is lowest for the beryllium and boron-like ions (e.g., \ion{O}{v}, \ion{O}{iv}), and it increases with atomic number ($Z$) along the sequence.
A striking feature is that for all ions DR becomes suppressed at densities  lower than typical solar atmosphere values ($10^9$~cm$^{-3}$), and so suppression is important to consider for studies of all solar features, particularly for $Z\le 10$.

\begin{deluxetable}{lccc}
\tablecaption{DR suppression data for O, Ne and Mg ions.\label{tbl.act}}
\tablehead{Ion & Log\,($T_{\rm max}$/K) & Log\,($N_{\rm e}$/cm$^{-3}$) & $S(T_{\rm max})$ }
\startdata
\ion{O}{iii} &     4.85 &     3.95 &   0.57 \\
\ion{O}{iv} &     5.10 &     3.82 &   0.56 \\
\ion{O}{v} &     5.35 &     3.22 &   0.53 \\
\ion{Ne}{iv} &     5.20 &     5.42 &   0.79 \\
\ion{Ne}{v} &     5.40 &     6.37 &   0.86 \\
\ion{Ne}{vi} &     5.55 &     5.76 &   0.80 \\
\ion{Ne}{vii} &     5.70 &     5.17 &   0.74 \\
\ion{Mg}{v} &     5.45 &     6.66 &   0.89 \\
\ion{Mg}{vi} &     5.65 &     7.45 &   0.95 \\
\ion{Mg}{vii} &     5.80 &     8.09 &   0.98 \\
\ion{Mg}{viii} &     5.90 &     7.41 &   0.95 \\

\enddata
\end{deluxetable}

\end{document}